\documentclass[12pt]{article}
\usepackage[pdftex]{graphicx}
\pdfoutput=1
\usepackage{latexsym}
\usepackage{amsmath,amssymb}

\newcommand\pubnumber{}
\newcommand\pubdate{\today}

\def\Title#1{\begin{center} {\Large #1 } \end{center}}
\def\Author#1{\begin{center}{ \sc #1} \end{center}}
\def\Address#1{\begin{center}{ \it #1} \end{center}}

\newcommand\pubblock{\rightline{\begin{tabular}{l} \pubnumber\\
         \pubdate  \end{tabular}}}
\newenvironment{Abstract}{\begin{quotation}  }{\end{quotation}}
\newenvironment{Presented}{\begin{quotation} \begin{center} 
             PRESENTED AT\end{center}\bigskip 
      \begin{center}\begin{large}}{\end{large}\end{center} \end{quotation}}
\def\Acknowledgements{\bigskip  \bigskip \begin{center} \begin{large}
             \bf ACKNOWLEDGEMENTS \end{large}\end{center}}

\def\beq{\begin{equation}}
\def\eeq#1{\label{#1}\end{equation}}
\def\eeqn{\end{equation}}

\def\beqa{\begin{eqnarray}}
\def\eeqa#1{\label{#1}\end{eqnarray}}
\def\eeqan{\end{eqnarray}}

\let\bar=\overbar

\def\Dslash{\not{\hbox{\kern-4pt $D$}}}
\def\dslash{\not{\hbox{\kern-2pt $\del$}}}

\def\msb{{\bar{\ssstyle M \kern -1pt S}}}

\begin{document}
\begin{titlepage}
\pubblock

\vfill
\Title{Characteristics of Non-Irradiated and Irradiated Double SOI Integration Type Sensor}
\vfill
\Author{Mari~ASANO$\mathrm{^{A}}$, Kazuhiko~HARA$\mathrm{^{A}}$, Daisuke~SEKIGAWA$\mathrm{^{A}}$, Shunsuke~HONDA$\mathrm{^{A}}$, Naoshi~TOBITA$\mathrm{^{A}}$, Yasuo~ARAI$\mathrm{^{B}}$, Toshinobu~MIYOSHI$\mathrm{^{B}}$, Ikuo~KURACHI$\mathrm{^{B}}$}
\Address{$\mathrm{^{A}}$University of Tsukuba, Tsukuba, Ibaraki Japan.\\
$\mathrm{^{B}}$High Energy Accelerator Research Organization(KEK), Tsukuba, Ibaraki, Japan.}
\vfill
\begin{Abstract}
We are developing monolithic pixel sensors based on a 0.2 $\mu$m fully-depleted Silicon-on-Insulator (SOI) technology for HEP experiment applications. The total ionizing dose (TID) effect is the major issue in the applications for hard radiation environments in HEP experiments. To compensate for TID damage, we have introduced a Double SOI structure which has a Middle Silicon layer (SOI2 layer) in addition. We studied the recovery from TID damage induced by $\mathrm{^{60}Co}~\gamma$'s and other characteristics of an Integration-type Double SOI sensor. The Double SOI sensor irradiated to 100 kGy showed a response for IR laser similar to of a non-irradiated sensor when we applied a negative voltage to the SOI2 layer. We conclude that the Double SOI sensor is radiation hard enough to be used in HEP experiments in harsh radiation environments such as at Bell I\hspace{-.1em}I or ILC.
\end{Abstract}
\vfill
\begin{Presented}
International Workshop on SOI Pixel Detector (SOIPIX2015), Tohoku University, Sendai, Japan, 3-6, June, 2015.
\end{Presented}
\vfill
\end{titlepage}
\def\thefootnote{\fnsymbol{footnote}}
\setcounter{footnote}{0}

\section{Introduction}
\begin{subsection}{SOI pixel sensor}
~~~We are developing monolithic pixel sensors for various applications \cite{TMiyoshi} based on a 0.2 $\mu$m fully-depleted SOI technology. The schematics of an SOI pixel sensor is shown in Figure \ref{fig:SOIPIX}\cite{SOIPIX}.
The sensors are processed by Lapis Semiconductor\cite{Lapis} on SOI wafers provided by SOITEC Co.\cite{SOITEC}. For use in hard radiation environment as in HEP experiments, TID effect is the major issue\cite{hondaTIPP}. 
\begin{figure}[!htbp]
\centering
\includegraphics[width=110mm]{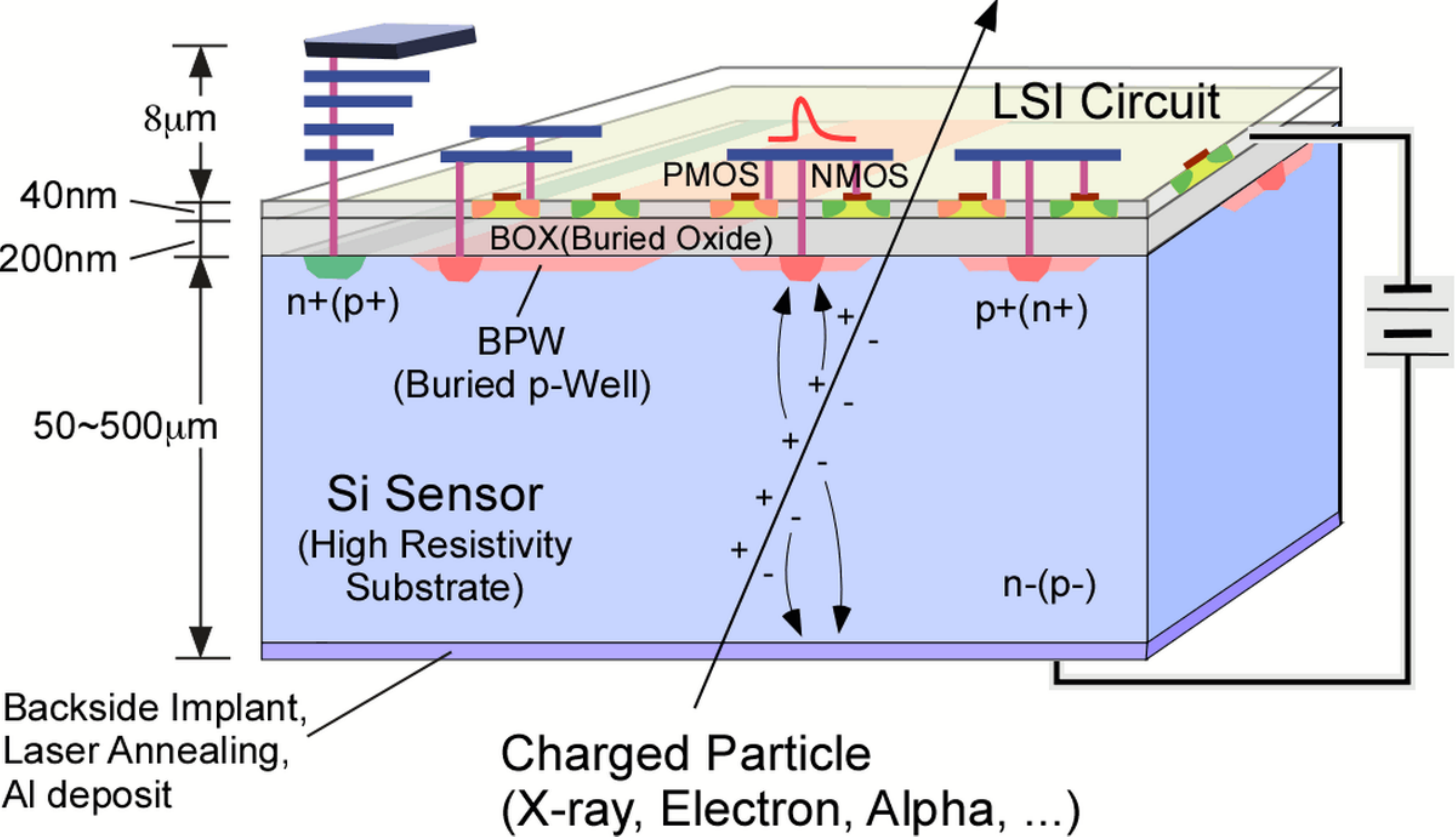}
\caption{Schematics of SOI pixel sensor.}
\label{fig:SOIPIX}
\end{figure}
\end{subsection}

\begin{subsection}{Double SOI}
~~~To compensate for TID damage, we introduced a Middle Silicon layer (SOI2 layer). By applying a negative voltage ($\mathrm{V_{SOI2}}$), the effects due to irradiation-induced holes trapped in oxide layers are to be cancelled (Figure \ref{fig:DSOI}).
\begin{figure}[!htbp]
\centering
\includegraphics[width=110mm]{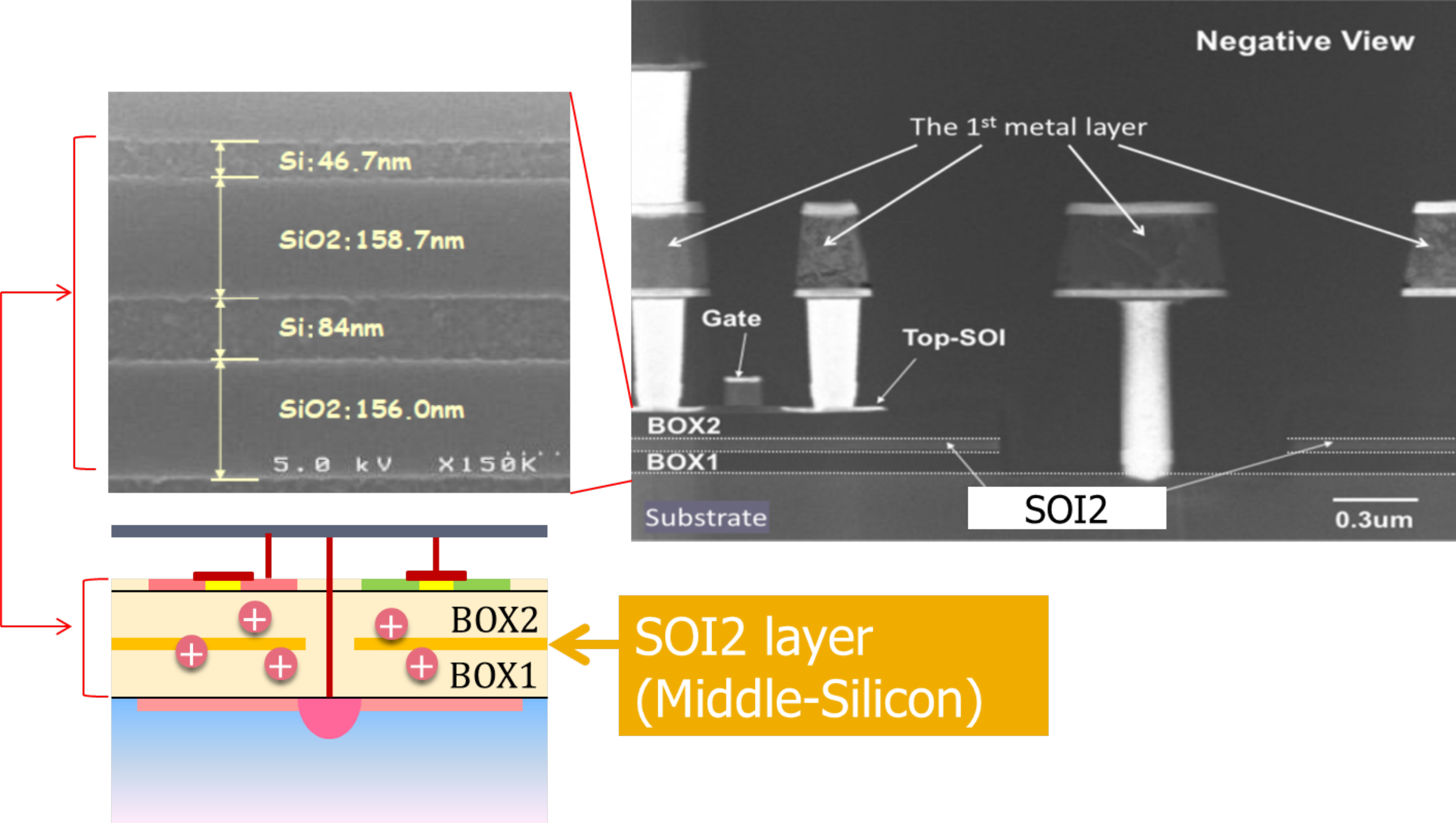}
\caption{TEM images (courtesy of Lapis Semiconductor) and schematics of the Double SOI structure.}
\label{fig:DSOI}
\end{figure}
\end{subsection}
\begin{subsection}{INTPIXh2}
~~~INTPIXh2 is an integration type SOI pixel sensor (see Table \ref{tab:INTh2Para} for main parameters). Figure \ref{fig:INTh2circuit} is the on-pixel circuit of INTPIXh2. The pixel architecture is shown in Figure \ref{fig:INTh2architecture}. The characteristics are compared between Double SOI and Single SOI types, which are fabricated in Double SOI wafer and Single SOI wafer, respectively. In the Double type, only the pixels in top-right sensor region (region-2) utilize the SOI2 layer (Figure \ref{fig:INTh2region}). We used them to study the recovery from TID damage induced by $\mathrm{^{60}Co}~\gamma$'s and to compare characteristics difference between Single SOI and Double SOI type sensors.
\begin{table}[!htbp]
\caption{INTPIXh2 parameters.}
\label{tab:INTh2Para}
\begin{center}
\begin{tabular}{l|c}\hline
\multicolumn{2}{c}{INTPIXh2 parameters}\\ \hline
 Pixel size & 18 $\mu$m $\square$ \\
Number of pixels & 280$\times$240\\
Overall chip size & 6mm $\square$ \\ \hline
\end{tabular}
\end{center}
\end{table}

\begin{figure}[htbp]
\begin{tabular}{ccc}
\begin{minipage}{0.475\hsize}
\begin{center}
\includegraphics[width=55mm]{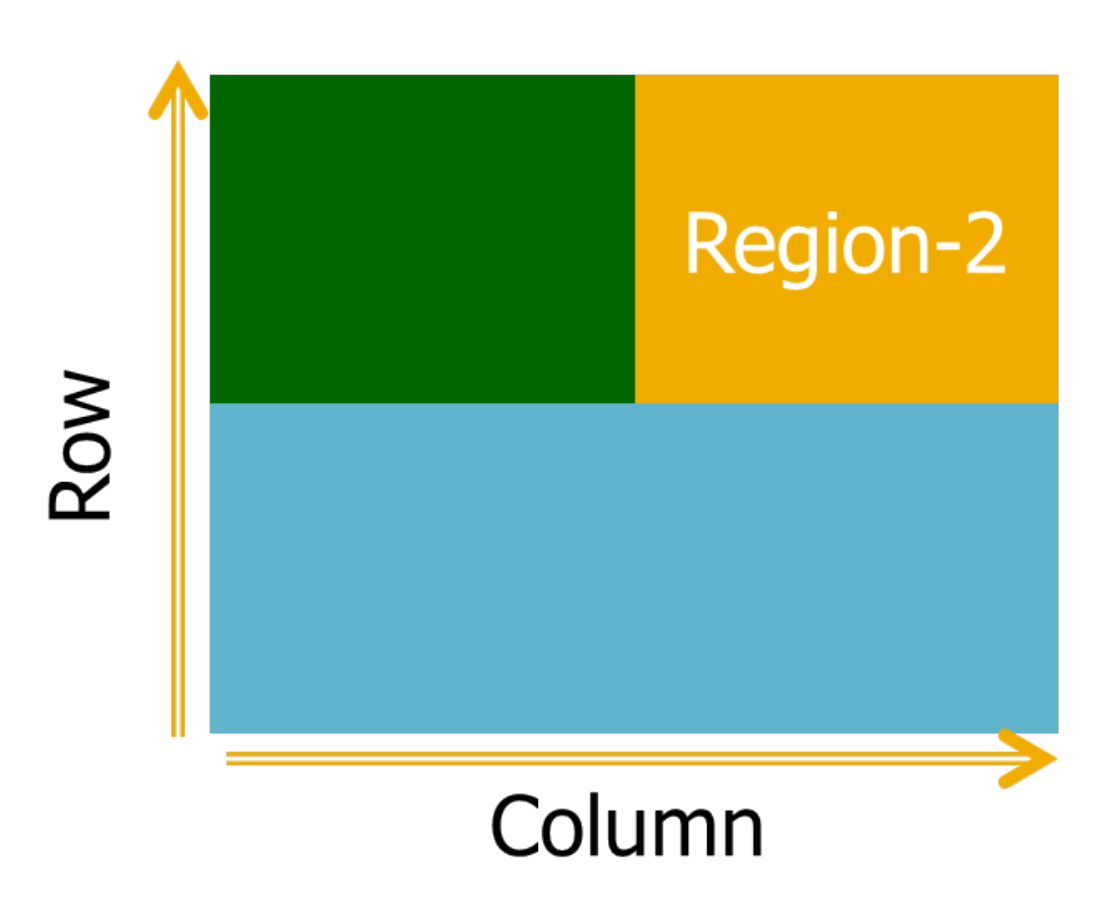}
\caption{The regions in INTPIXh2.\protect\\ Only the pixels in Region-2 utilize the SOI2.}
\label{fig:INTh2region}
\end{center}
\end{minipage}
\begin{minipage}{0.05\hsize}~~
\end{minipage}
\begin{minipage}{0.475\hsize}
\begin{center}
\includegraphics[width=80mm]{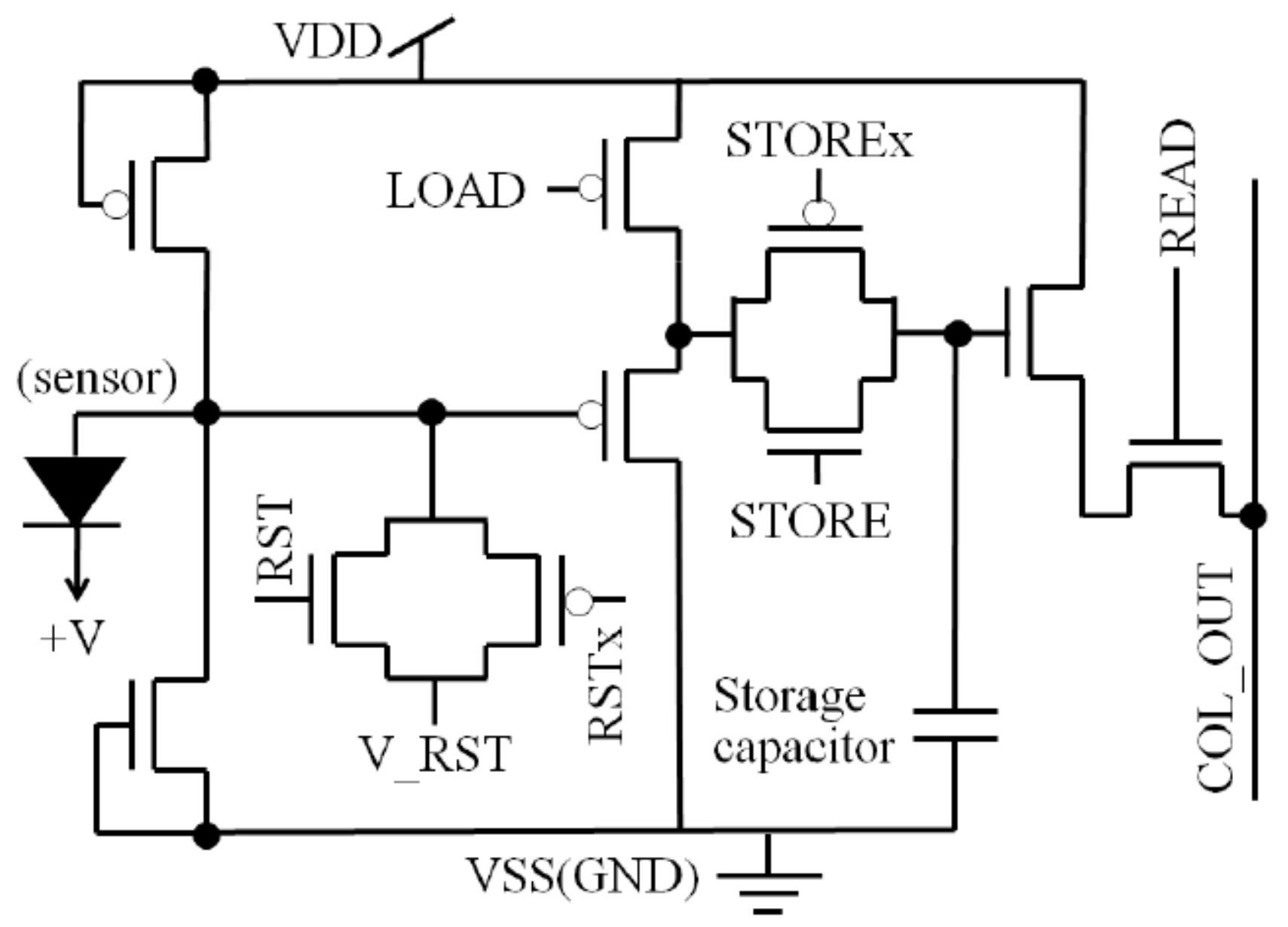}
\caption{On-pixel circuit of INTPIXh2.}
\label{fig:INTh2circuit}
\end{center}
\end{minipage}
\end{tabular}
\end{figure} 

\begin{figure}[htbp]
\centering
\includegraphics[width=70mm]{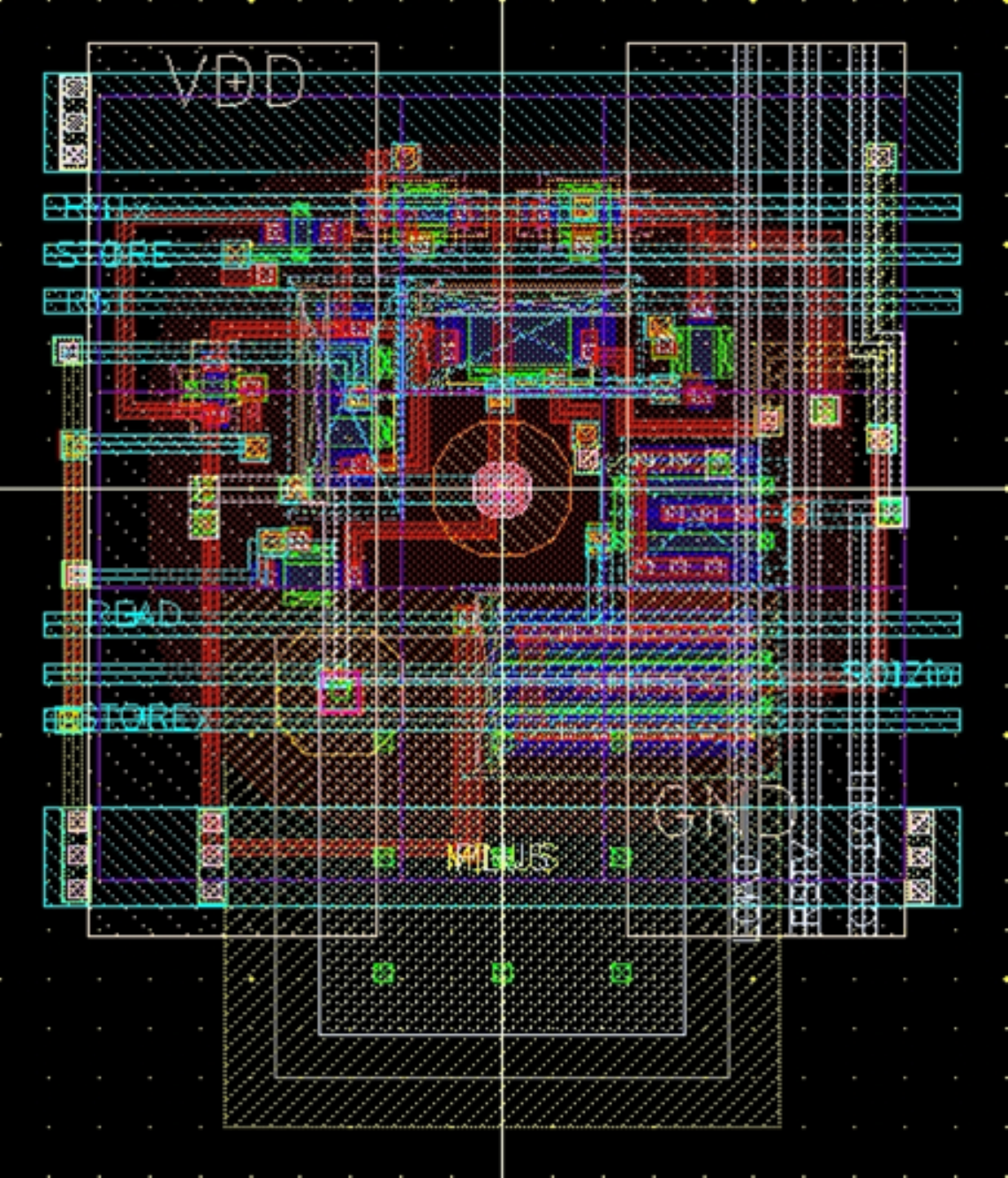}
\caption{Pixel architecture of INTPIXh2.}
\label{fig:INTh2architecture}
\end{figure}
\end{subsection}

\section{Response after 100 kGy irradiation}
\begin{figure}[htbp]
\centering
\includegraphics[width=130mm]{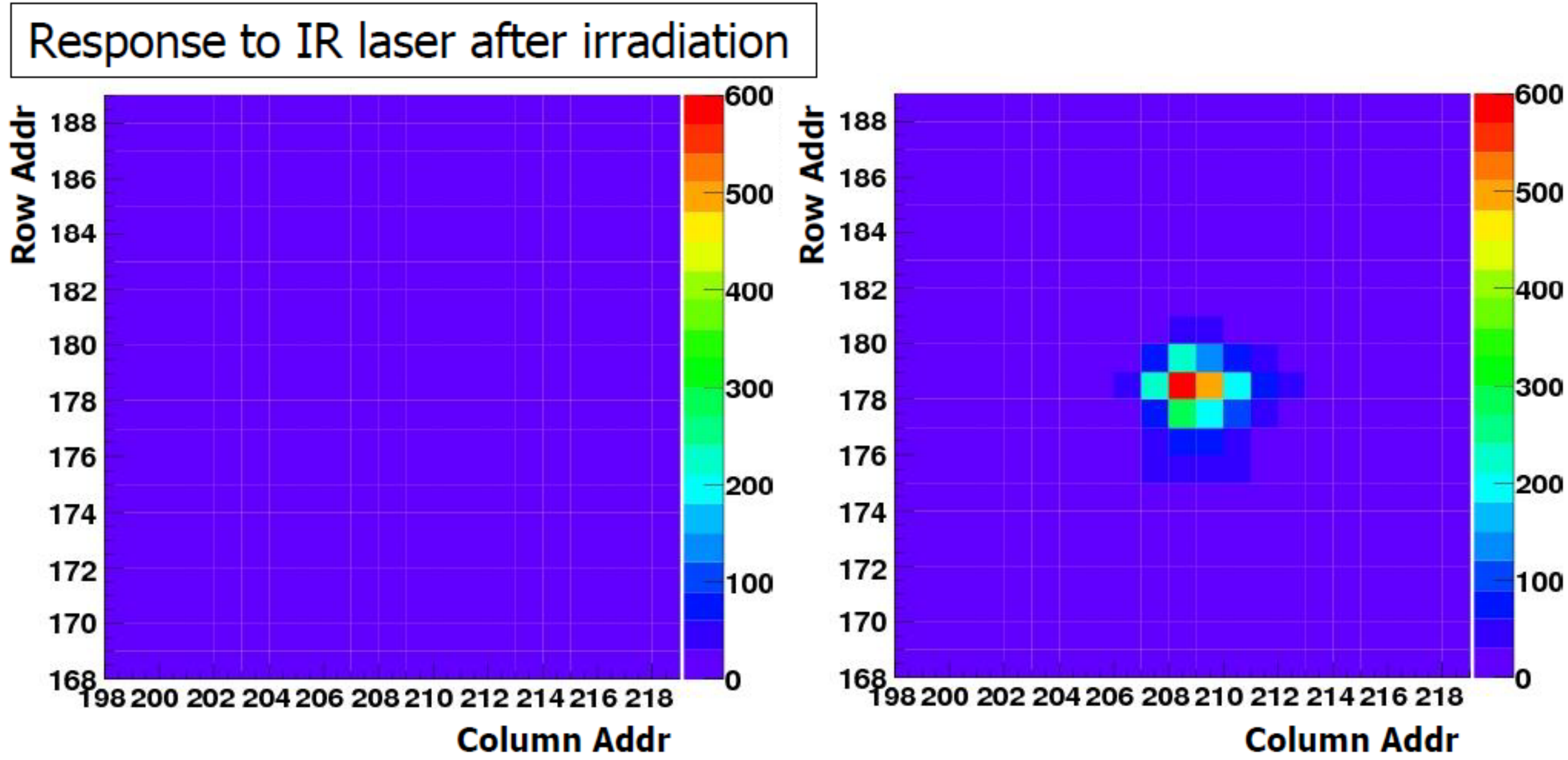}
\caption{Response to IR laser after 100 kGy irradiation. (left) $\mathrm{V_{SOI2}=0.0~V}$, (right) $\mathrm{V_{SOI2}=-10.0~V}$ }.
\label{fig:IRresponse}
\end{figure}

\begin{figure}[htbp]
\centering
\includegraphics[width=100mm]{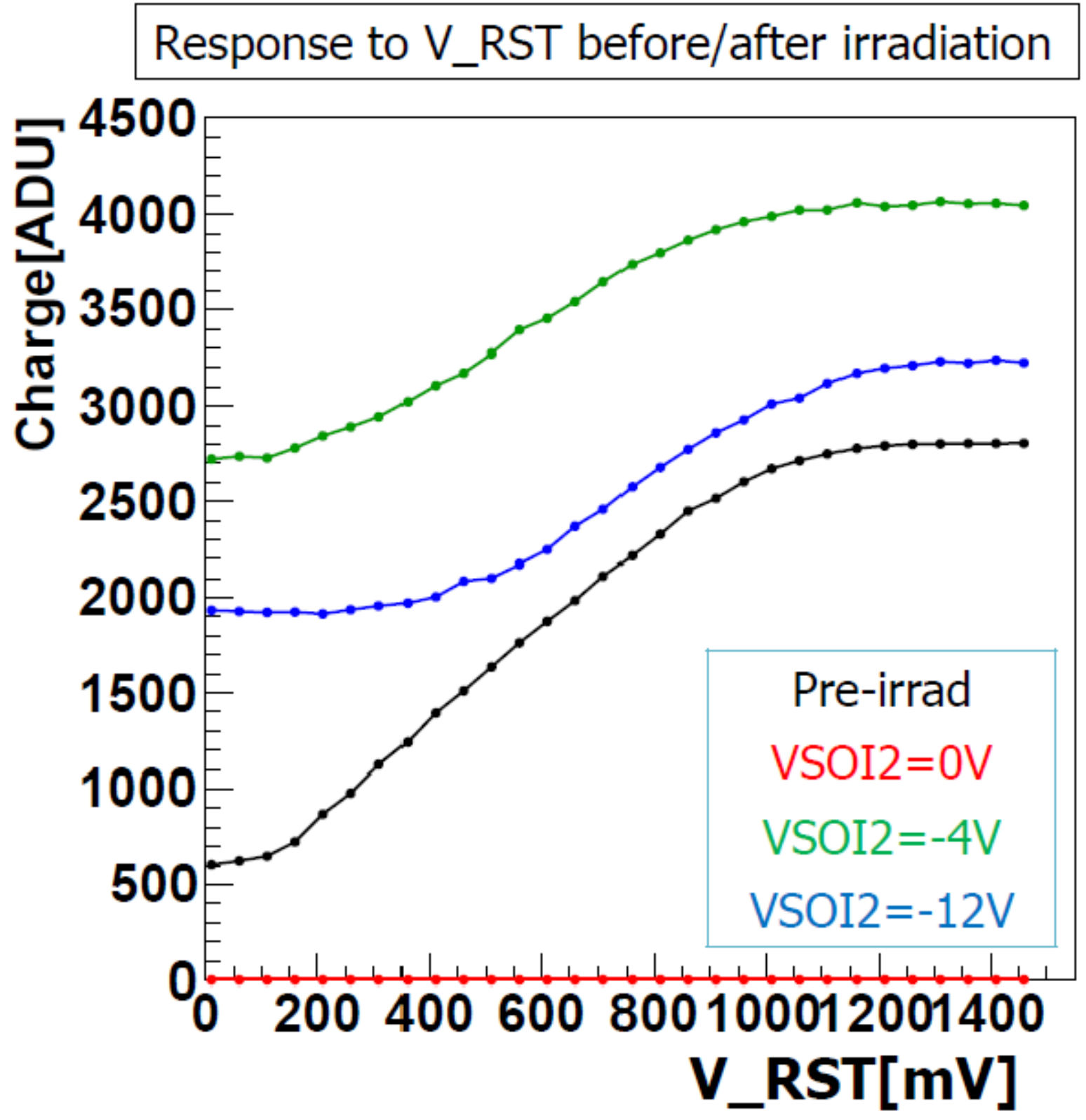}
\caption{Response to $\mathrm{V_{RST}}$ before and after 100 kGy irradiation.}
\label{fig:RSTVresponse}
\end{figure}

~~~Response of a 100 kGy irradiated sensor (300 $\mu$m thick, 100 V applied as reverse bias) to IR laser was clearly confirmed at $\mathrm{V_{SOI2}=-10~V}$ while it was not at all at $\mathrm{V_{SOI2}=0~ V}$ (see Figure \ref{fig:IRresponse}). This demonstrates that applying a negative voltage to the SOI2 layer recovers the response of irradiated sensor to the level of non-irradiated one.\\
~~~We studied the functionality of the circuit by measuring the output linearity w.r.t. the reset voltage amplitude ($\mathrm{V_{RST}}$). 
Figure \ref{fig:RSTVresponse} plots the results for the 100 kGy irradiated sensor (100 $\mu$m thick, 100 V applied as reverse voltage), showing the recovery of the response by applying negative voltages to $\mathrm{V_{SOI2}}$ (green, blue). There was no response seen at $\mathrm{V_{SOI2}=0 V}$ (red). The response curve of a non-irradiated sensor is also plotted for comparison (black). The shifts in operation conditions, mainly thresholds, of the FETs induced by irradiation have not been completely recovered because a single common $\mathrm{V_{SOI2}}$ is applicable in this sensor design for various types of FETs employed. However, the response region has been clearly re-established by $\mathrm{V_{SOI2}}$ with the dynamic range being reduced partly. 

\section{Signal speed}
\begin{figure}[htbp]
\begin{tabular}{ccc}
\begin{minipage}{0.475\hsize}
\begin{center}
\includegraphics[width=70mm]{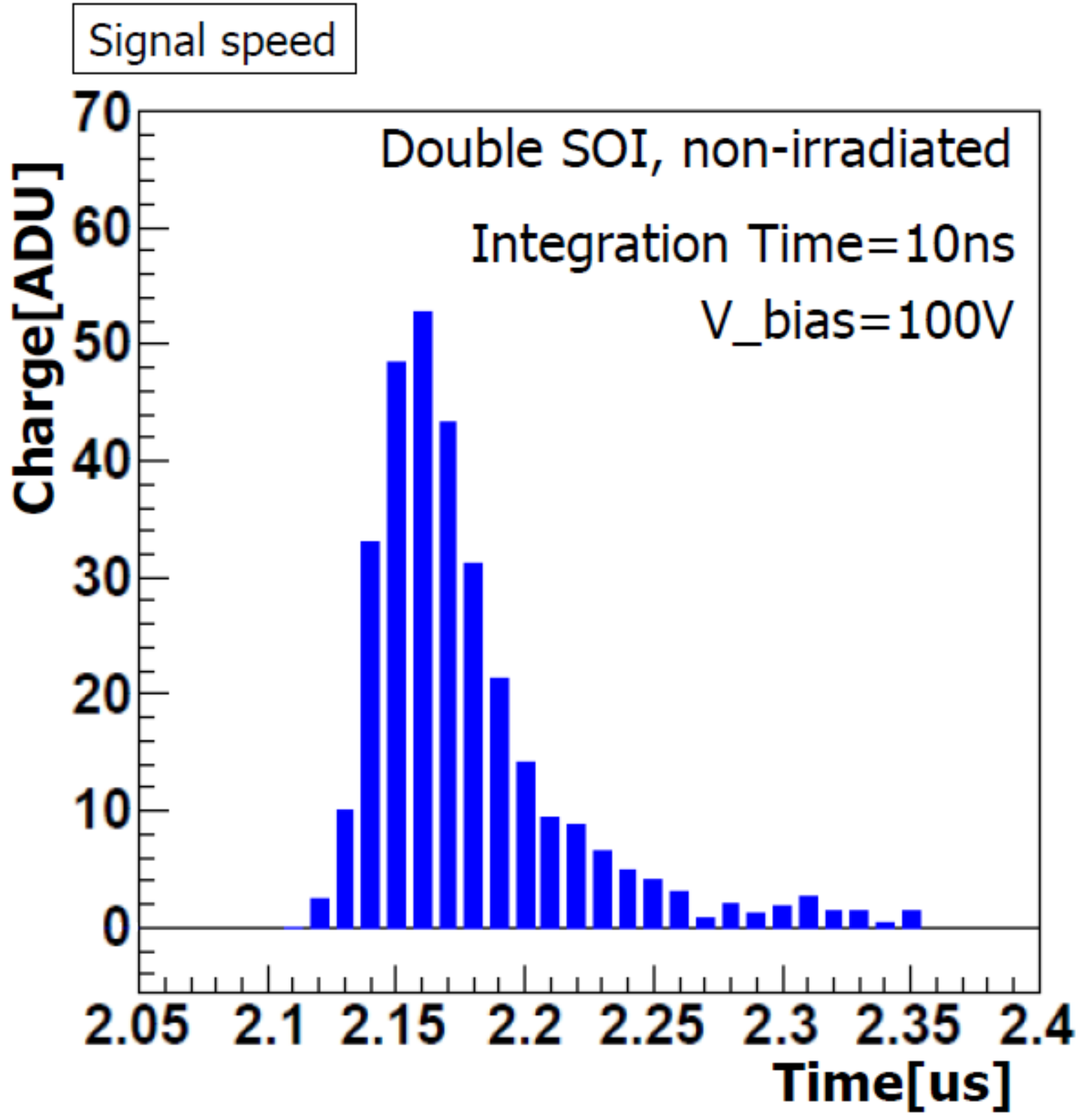}
\caption{Signal shape of INTPIXh2.}
\label{fig:SignalSpeed}
\end{center}
\end{minipage}
\begin{minipage}{0.05\hsize}~~
\end{minipage}
\begin{minipage}{0.475\hsize}
\begin{center}
\includegraphics[width=70mm]{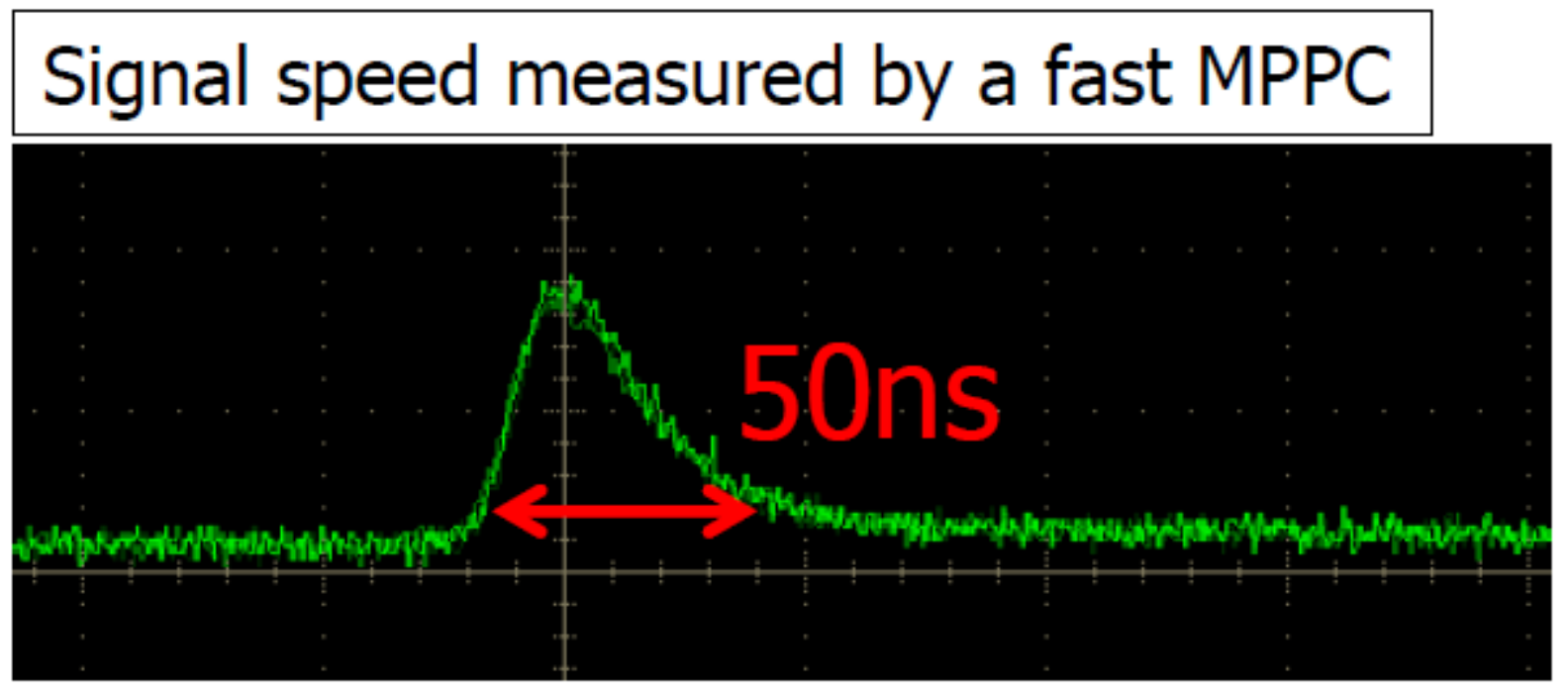}
\caption{Injected laser pulse shape measured with a fast MPPC.}
\label{fig:SignalSpeedMPPC}
\end{center}
\end{minipage}
\end{tabular}
\end{figure}

~~~We evaluated the signal time structure using a pulsed IR laser (wavelength 1064nm, signal width 50 ns, jitter ~5 ns). Figure \ref{fig:SignalSpeedMPPC} shows the IR laser signal shape measured by a fast MPPC. We measured the output charge from INTPIXh2 as a function of delay time with a fixed integration time of 10~ns. By varying the delay time of integration start w.r.t. the IR laser trigger, we derived the information how fast the signal of INTPIXh2 is. The measured signal width is about 150~ns with a fast peaking time of 40 ns (Figure \ref{fig:SignalSpeed}).

\section{Cross talk}
\begin{figure}[htbp]
\centering
\includegraphics[width=120mm]{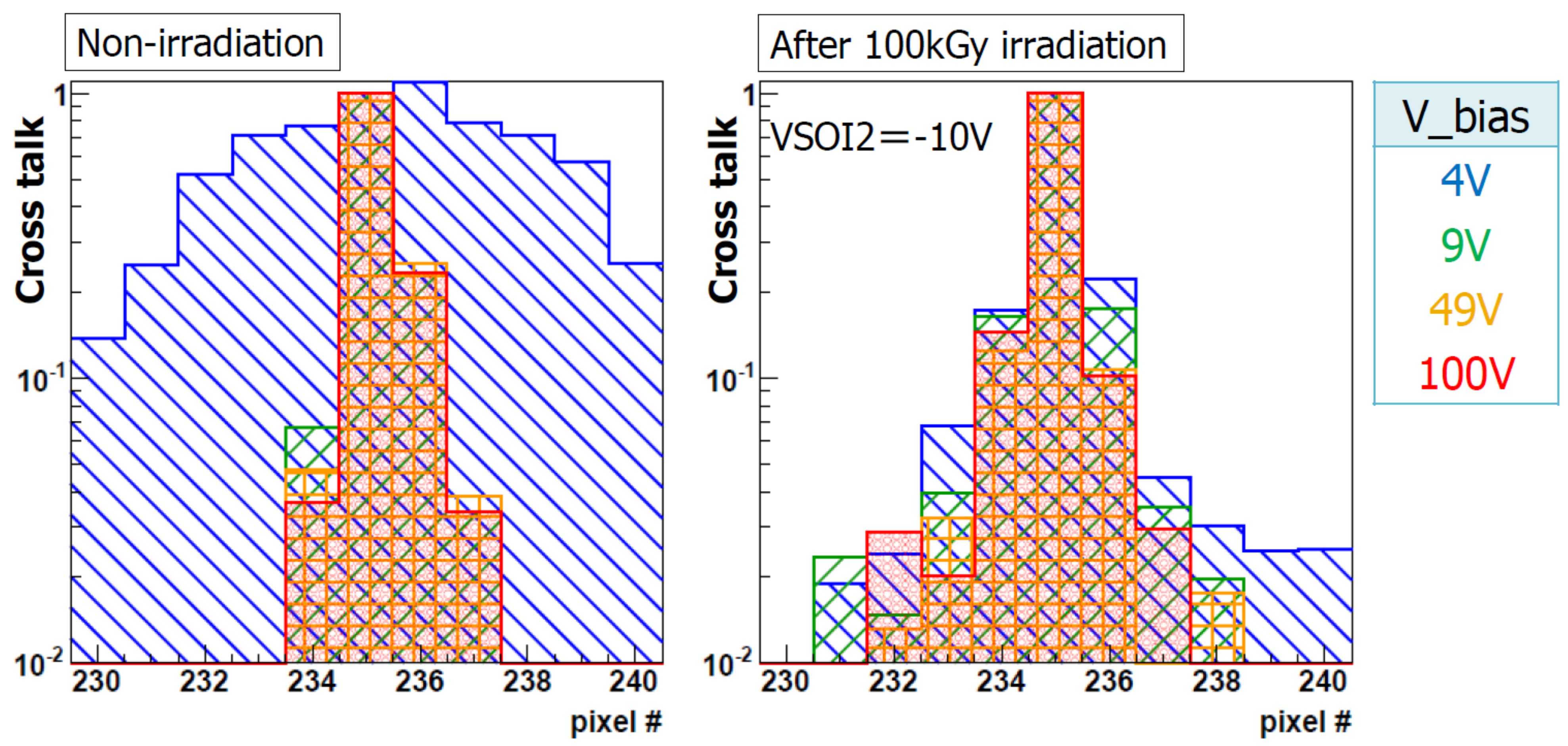}
\caption{Cross talk of INTPIXh2 before and after 100 kGy irradiation. \protect\\The collimated laser was spotted at pixel-235, and relative pulse heights in neighbouring pixels are plotted for typical bias voltages. }
\label{fig:xtalk}
\end{figure}

~~~We measured the output signals from a pixel in which IR laser was injected and from the surrounding pixels (100 $\mu$m thick n-bulk sensor). The laser spot size was set to 4 $\mu$m square in order to inject the laser to one pixel only. We define the cross talk as the ratio of output signal height from a pixel to that from the pixel where laser was injected. The cross talk at low reverse bias voltage was substantial before irradiation (Figure\ref{fig:xtalk}). However, the cross talk has decreased after irradiation. This is understood that the electrical isolation between pixels is improved by irradiation by accumulating electrons between p-type pixel nodes. The cross talk is found not to degrade by irradiation.

\section{Response to MIP}
\begin{figure}[htbp]
\centering
\includegraphics[width=150mm]{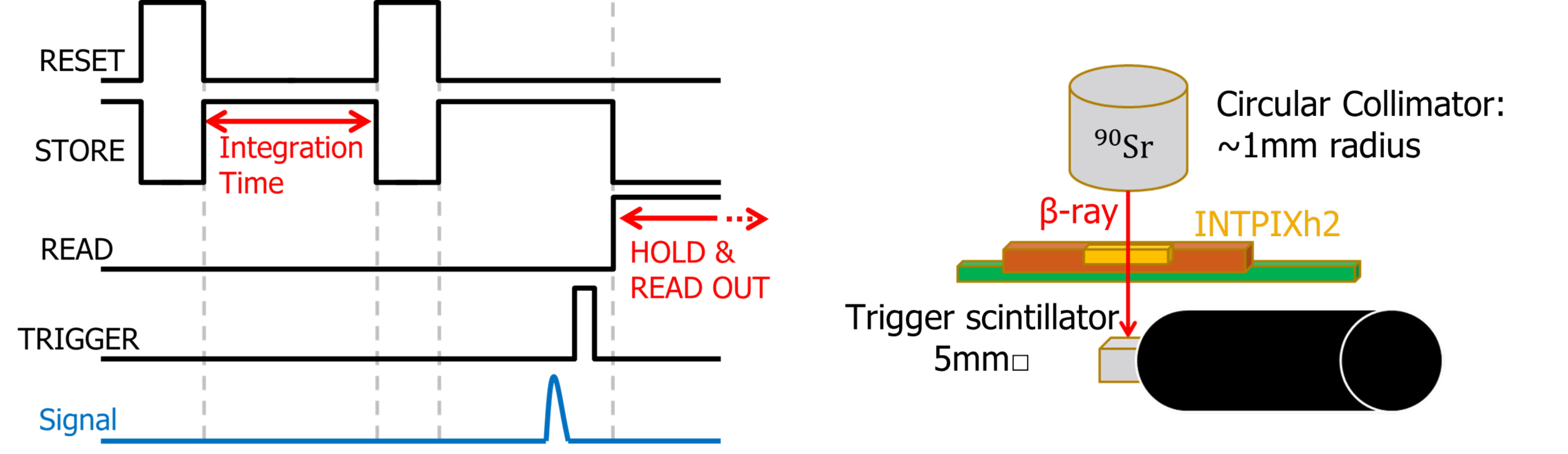}
\caption{(left) Trigger timing chart in the mip measurement. On receiving a trigger, the next reset is disabled and readout sequence starts. (right) Geometrical illustration of the setup.}
\label{fig:triggersystem}
\end{figure}

\begin{figure}[htbp]
\begin{tabular}{ccc}
\begin{minipage}{0.475\hsize}
\begin{center}
\includegraphics[width=75mm]{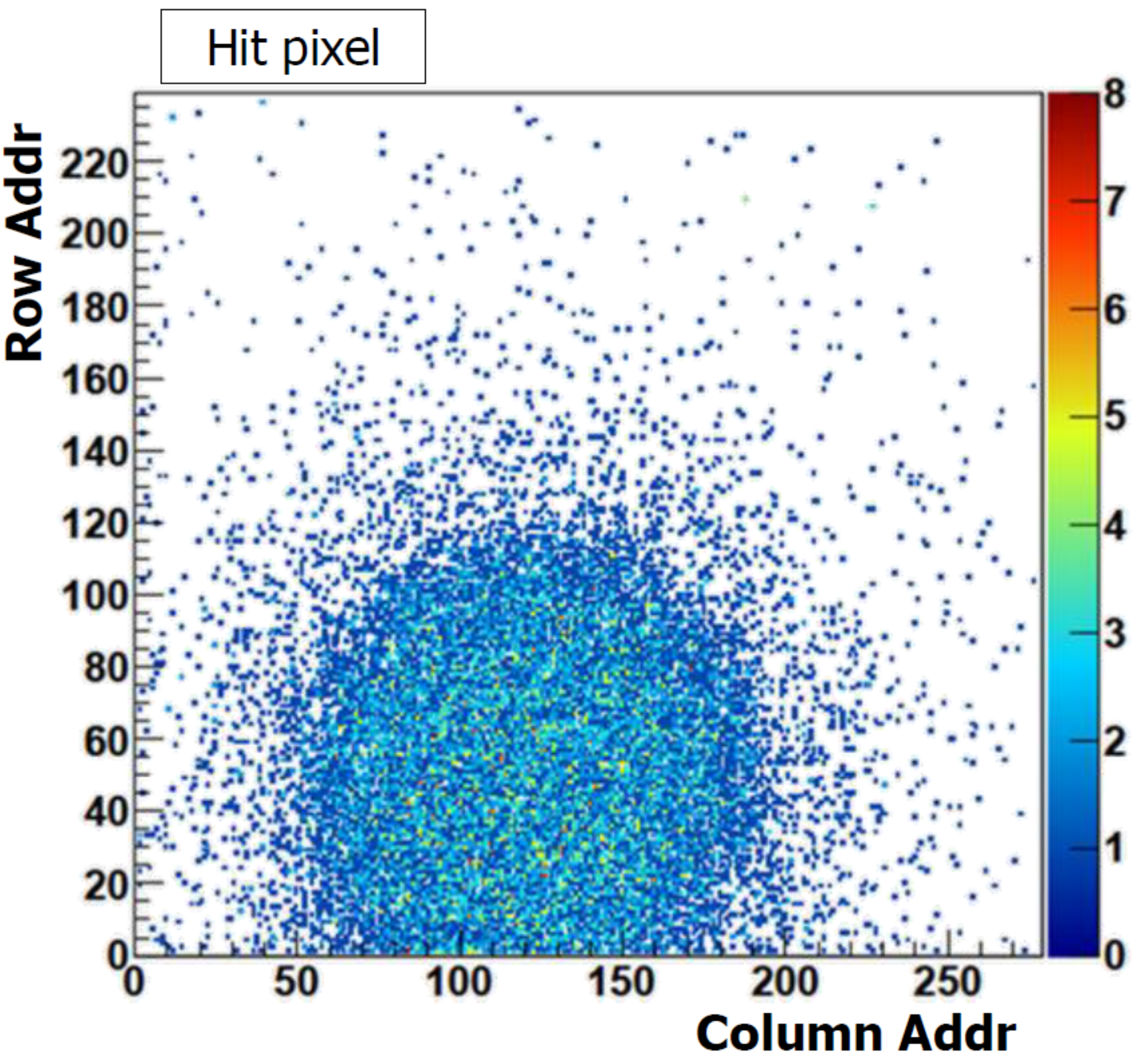}
\caption{Hit pixel distribution obtained in a single SOI sensor. The distribution reflects the shape of the $\beta$-ray collimator.}
\label{fig:HitSingle}
\end{center}
\end{minipage}
\begin{minipage}{0.05\hsize}~~
\end{minipage}
\begin{minipage}{0.475\hsize}
\begin{center}
\includegraphics[width=75mm]{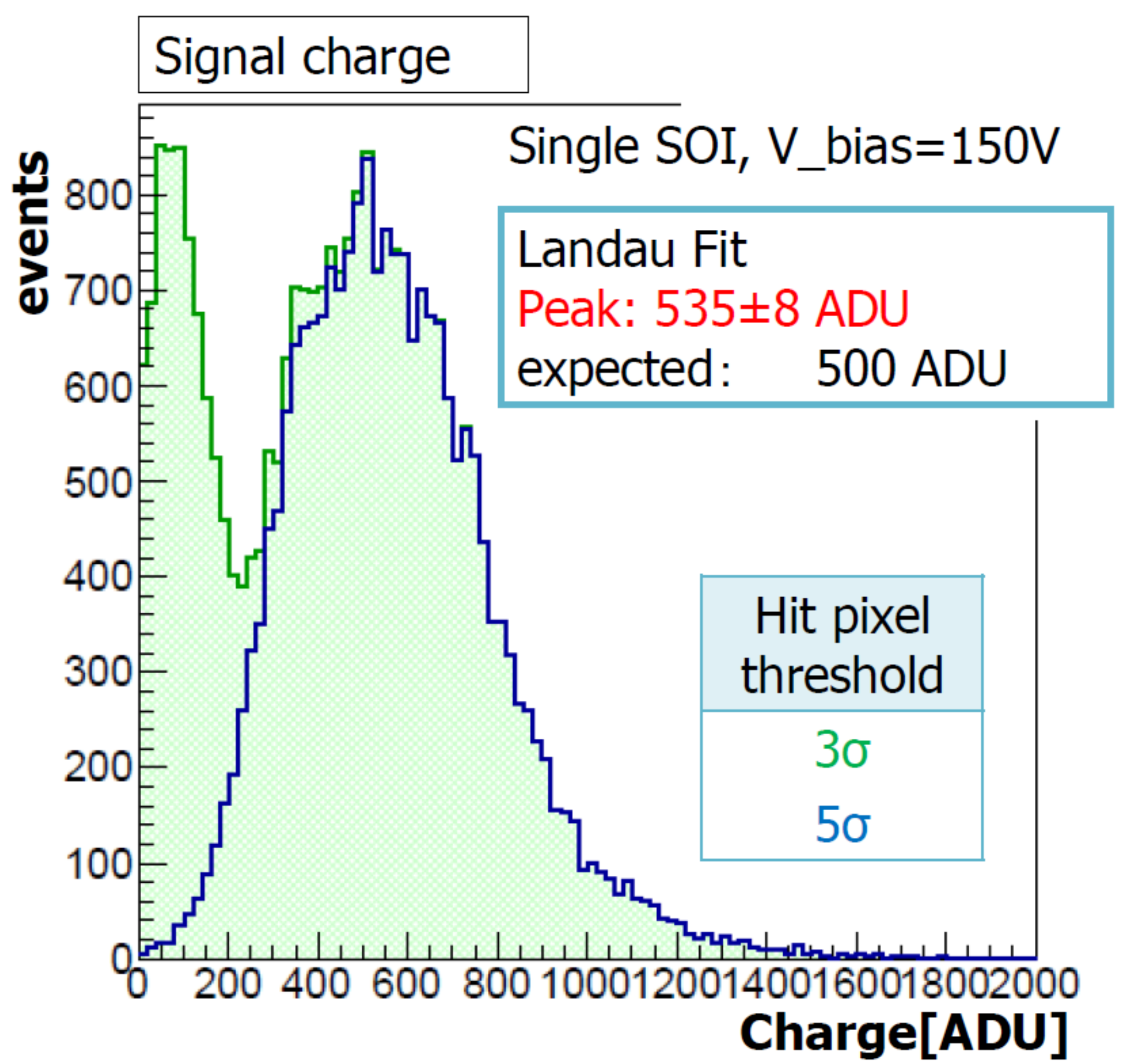}
\caption{Cluster charge distribution obtained in a single SOI sensor. The distributions are for two hit thresholds of 3$\sigma$ and 5$\sigma$ where $\sigma$ is the noise width calculated pixel by pixel.}
\label{fig:SignalSingle}
\end{center}
\end{minipage}
\end{tabular}
\end{figure}

\begin{figure}[htbp]
\begin{tabular}{ccc}
\begin{minipage}{0.475\hsize}
\begin{center}
\includegraphics[width=75mm]{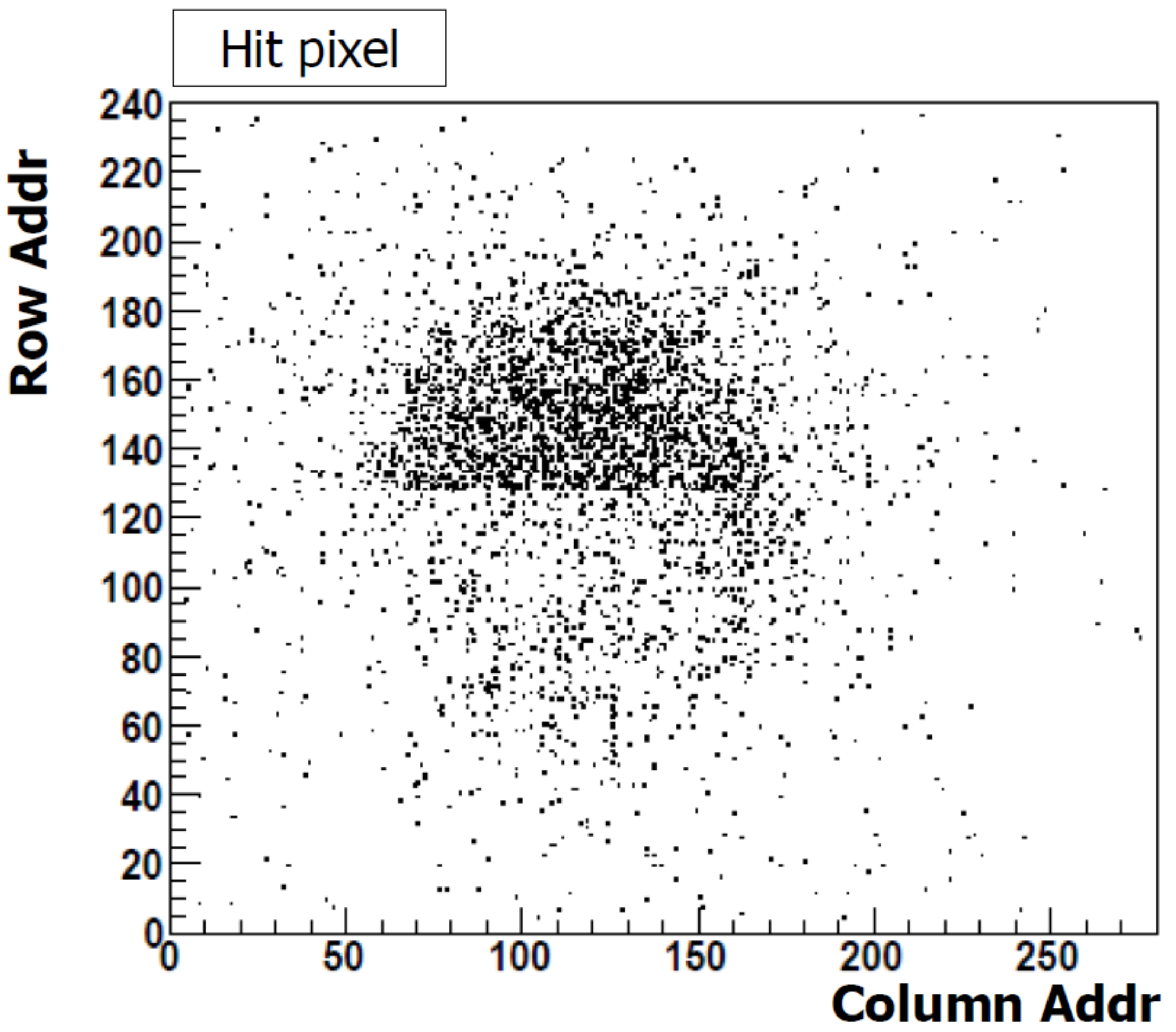}
\caption{Hit pixel distribution obtained in a double SOI sensor. Since the gains in Regions are different, the distribution is not symmetric. The double SOI region is top-right.}
\label{fig:HitDouble}
\end{center}
\end{minipage}
\begin{minipage}{0.05\hsize}~~
\end{minipage}
\begin{minipage}{0.475\hsize}
\begin{center}
\includegraphics[width=75mm]{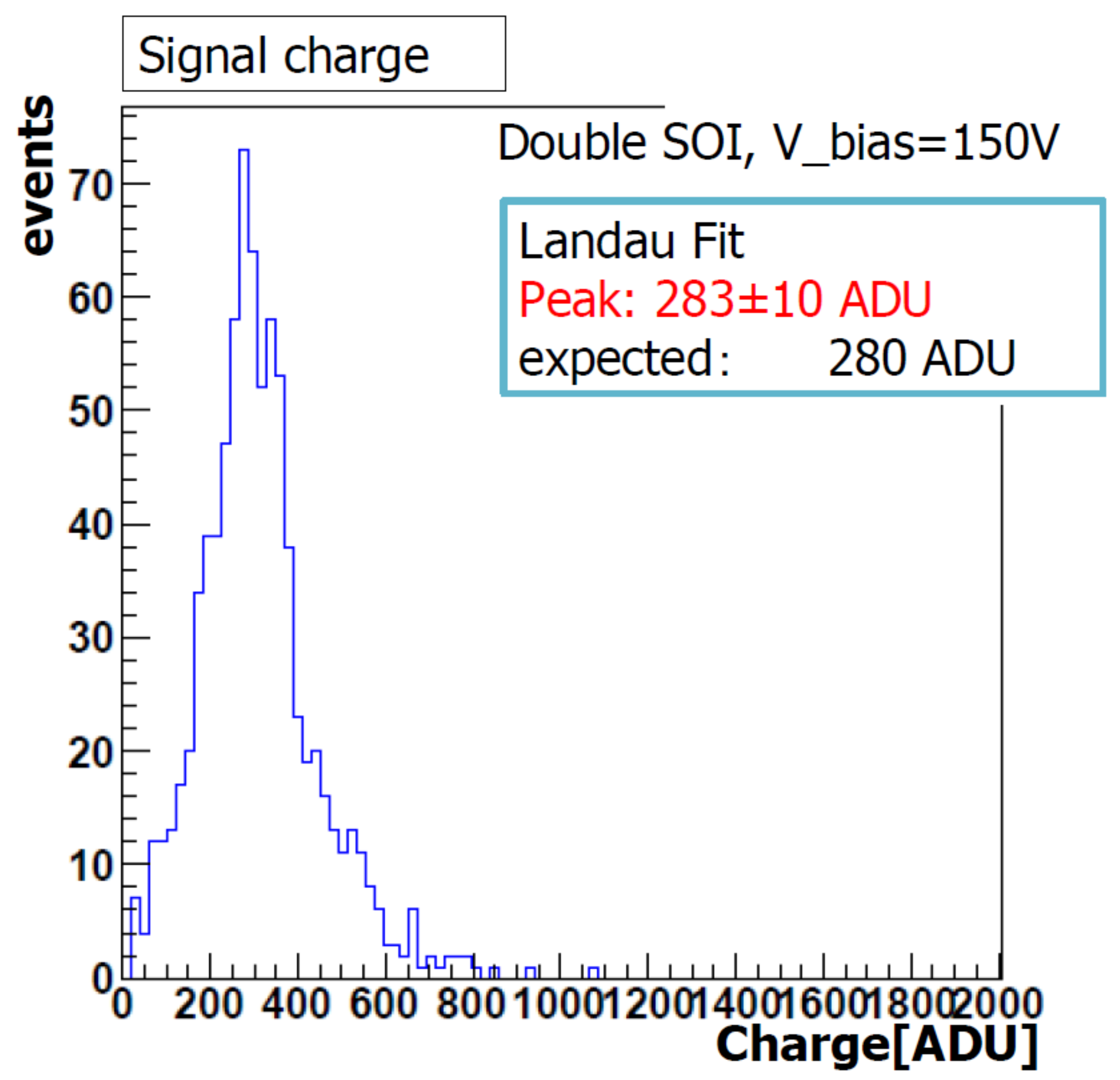}
\caption{Cluster charge distribution obtained by pixels in double SOI region.}
\label{fig:SignalDouble}
\end{center}
\end{minipage}
\end{tabular}
\end{figure}

~~~The response to penetrating $\mathrm{^{90}Sr}~\beta$-rays was measured with INTPIXh2 Single and Double sensors. The $\beta$-ray source was placed on top and penetrated $\beta$ was triggered by a plastic scintillator. The trigger system timing chart and geometry illustration are shown in Figure \ref{fig:triggersystem}. Table \ref{tab:MIPPara} summarizes the parameters of INTPIXh2 set for the measurement. The $\beta$-ray signal was reconstructed in a 5$\times$5 pixel cluster with the seed centred at the pixel which has the maximum response in an event. Hit pixel distributions of the seeds in single-type and double-type sensors are shown in Figure \ref{fig:HitSingle} and Figure \ref{fig:HitDouble}, respectively. The response of a single-type sensor showed a clear mip peak separated well from the pedestal with the peak value consistent with the calculation using a depletion thickness of 315 $\mu$m (Figure \ref{fig:SignalSingle}). The response of a double-type sensor showed a mip peak with the peak as calculated using a 190 $\mu$m thickness (Figure \ref{fig:SignalDouble}).

\begin{table}[htbp]
\caption{Parameters in the MIP response measurement.}
\label{tab:MIPPara}
\begin{center}
\begin{tabular}{l|c}\hline
\multicolumn{2}{c}{INTPIXh2, single and double}\\ \hline
Integration Time  & 1.00 $\mu$s \\
$\mathrm{V_{RST}}$ & 400 mV \\
$\mathrm{V_{bias}}$ & 150 V \\ \hline
\end{tabular}
\end{center}
\end{table}

\section{Summary}
~~~We have introduced a double SOI structure for the purpose of compensating for TID damage. Characterization results obtained with non-irradiated and 100 kGy irradiated sensors conclude that the response from irradiated sensor is recoverable by applying negative voltage to the SOI2 layer. The double SOI sensor is radiation hard enough to be used in HEP experiments in hard radiation environment such as at Bell II or ILC.

\Acknowledgements
The authors are grateful for fruitful collaboration with the Lapis Semiconductor Co. Ltd. The double SOI wafers have been realized through their excellence. This work was supported by JSPS KAKENHI Grand Number 25109006, by KEK Detector
Technology Project and also by VLSI Design and Education Center (VDEC), The University of
Tokyo, with the collaboration of the Cadence Corporation and Mentor Graphics Corporation.


\begin{thebibliography}{99}


\bibitem{TMiyoshi}
T. Miyoshi et al., gMonolithic pixel detectors with 0.2 $\mu$m FD-SOI pixel process technology,h Nucl. Instr. Meth. A732 (2013) 530.

\bibitem{SOIPIX}
SOIPIX collaboration group: http://rd.kek.jp/project/soi/.

\bibitem{Lapis}
Lapis Semiconductor Co., Ltd.: http://www.lapis-semi.com/en/.

\bibitem{SOITEC}
SOITEC: http://www.soitec.com/en/technologies/smart-cut/.

\bibitem{hondaTIPP}
S. Honda et al., g Total Ionization Damage Compensations in Double Silicon-on-
Insulator Pixel Sensors h, Proceedings of Sciences (TIPP2014), 039, 2014.


\end{thebibliography}
\end{document}